\documentclass[pra,aps,showpacs,floats,twocolumn]{revtex4}
\usepackage{graphicx}
\usepackage{hyperref}
\usepackage{cleveref}

\begin{document}
\title{Molecular Kondo effect in flat-band lattices}
\author{Minh-Tien Tran$^{1,2,3}$ and Thuy Thi Nguyen$^2$}
\affiliation{$^1$Graduate University of Science and Technology, Vietnam Academy of Science and Technology,
Hanoi 100000, Vietnam. \\
$^2$Institute of Physics, Vietnam Academy of Science and Technology,
Hanoi 100000, Vietnam.\\
$^3$Center for Theoretical Physics of Complex Systems, Institute for Basic Science, Daejeon 34126, Korea.}

\begin{abstract}
The Kondo effect of a single magnetic impurity embedded in the Lieb lattice is studied by the numerical renormalization group.  When the band flatness is present in the local density of states at the impurity site, it quenches the participation of all dispersive electrons in the Kondo singlet formation, and reduces the many-body Kondo problem to a two-electron molecular Kondo problem. A quantum entanglement of two spins, which is the two-electron molecular analog of the many-body Kondo singlet, is stable at low temperature, and the impurity contributions to thermodynamical and dynamical quantities are qualitatively different from that obtained in the many-body Kondo effect. The conditions for existence of the molecular Kondo effect in narrow band systems are also presented.
\end{abstract}


\maketitle

\section{Introduction}

For a half century, the Kondo effect has attracted great scientific interest due to its essential relation to
a number of spectacular low temperature phenomena \cite{Hewson}. It is known to be responsible for the essential behaviors of heavy fermions, mixed valent and intermetallic alloys, Kondo insulators, transport and magnetic properties of quantum dots,... etc. The Kondo effect is inherently a strongly correlated many-body ground state that contains a quantum entanglement between a localized fermion and itinerant fermions. In this quantum entanglement many itinerant fermions in the Fermi sea are involved in order to quench the magnetic moment of the localized fermion and they together form the so-called Kondo singlet state. A molecular analog of the Kondo singlet state, where only a single fermion is involved in forming the spin singlet state with the localized fermion, was proposed \cite{Fuldebook,Fulde}. It was suggested to be the underlying physics of the Kondo effect experimentally observed in single molecules \cite{Fuldebook,Fulde,Booth}. The molecular Kondo effect contains key ingredients of the many-body Kondo effect in bulk. At sufficient low temperature, it separates the spin and charge low-lying excitations. The ground state is the spin singlet like the many-body Kondo singlet state. As temperature increases, the next higher energy state, the spin triplet, populates and forms a full magnetic moment \cite{Fuldebook}. These properties are exactly the main features of the many-body Kondo effect in bulk. However,
the molecular Kondo effect is also distinct from the many-body one by the absence of the Kondo resonance, which results from the coherently cotunneling of electrons between the Fermi sea and the magnetic impurity. Although, the molecular Kondo singlet is the underlying physics of the many-body Kondo effect in bulk, it cannot be isolated from the many-body Kondo singlet state.

In this paper we show the molecular Kondo effect can be observed in flat-band and narrow-band lattices. Unlike the many-body Kondo effect, the molecular Kondo singlet state is only a two electron spin state, which yields essentially an entanglement of two spin qubits.
This paves the way for implementing a two electron spin entanglement in solids.
The entangled qubit pair is a key element for conveying quantum information through a quantum device in quantum computation \cite{qi}.
Although the entanglement of two spin qubits has extensively been investigated, it has been demonstrated only in quantum dots \cite{Shulman} and in molecules \cite{Garlatti}. In the flat-band lattices, the two-electron Kondo singlet is isolated, leaving all other electrons in the Fermi sea irrelevant. It is essentially the entanglement of two spin qubits in solids.
We will investigate the Kondo problem in the Lieb lattice, the electron structure of which features a band flatness\cite{Lieb,Tasakireview,Tien}. We calculate the thermodynamical and dynamical quantities by the numerical renormalization group (NRG) \cite{Wilson1,Wilson2,Wilson3,Bulla}. The NRG was originally constructed for solving the Kondo problem in normal metals. In principle, it is an exact tool for solving the Kondo problem in any systems.
We find that the band flatness quenches the participation of all dispersive electrons in the Kondo singlet formation, and reduces the many-body Kondo problem to a two-electron Kondo problem. The Kondo impurity in the flat band lattices yields the opposite limit of the one in graphene or pseudogap systems \cite{Vojta},  since the density of states (DOS) at the flat band is infinite. In the molecular Kondo effect the impurity contributions to the thermodynamical and dynamical quantities are qualitatively different from that of the many-body Kondo effect in normal metals \cite{Hewson}, pseudogap systems \cite{Vojta}, Kondo boxes \cite{Thimm}, and quantum dots \cite{Glazman}. This constitutes a novel regime of the Kondo problem in the presence of a band flatness.
We also show that the molecular Kondo effect can also exist in narrow-band lattices, providing the band width is narrow enough.

The present paper is organized as follows. In Sec. II we present the Anderson impurity model on the Lieb lattice. In this section we also describe the NRG for the flat-band lattice and present its results for the Kondo problem.
Finally, the conclusion is presented in Sec. III.

\section{Anderson impurity model in the Lieb lattice and its numerical renormalization group results}

\begin{figure}[b]
\centering
  \begin{tabular}{cc}
    \includegraphics[width=0.18\textwidth]{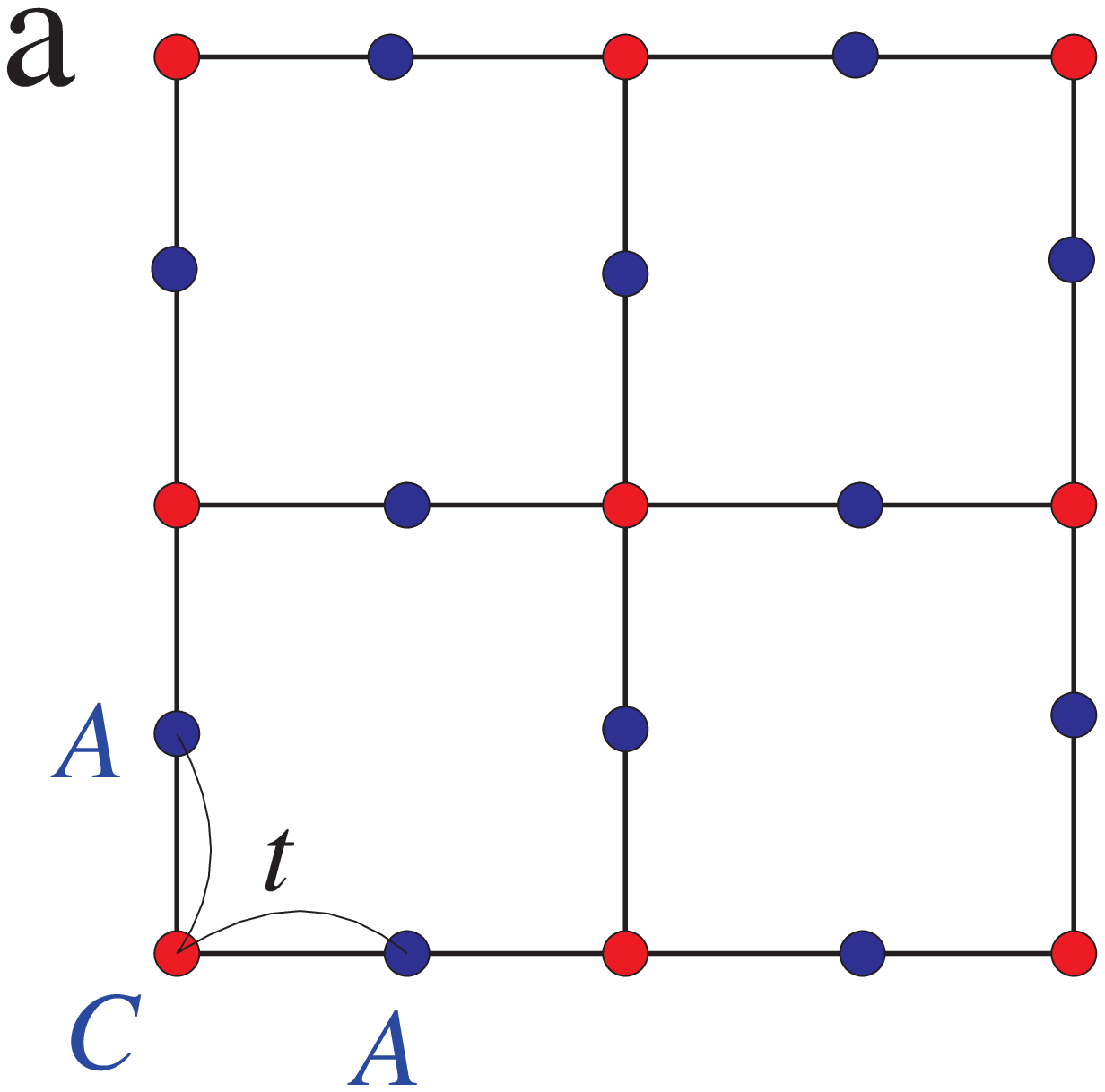} &
    \includegraphics[width=0.24\textwidth]{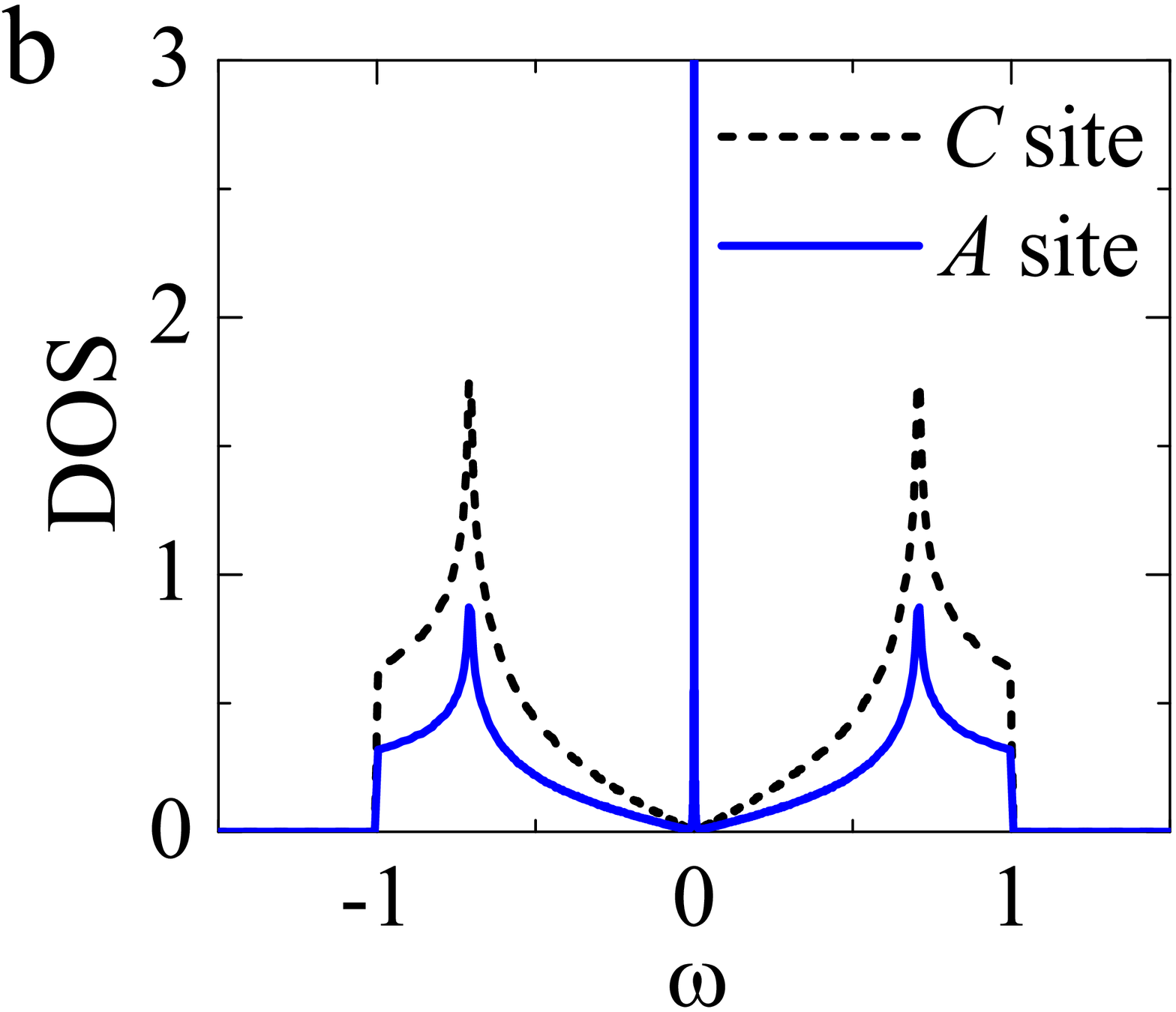} \\
    \includegraphics[width=0.24\textwidth]{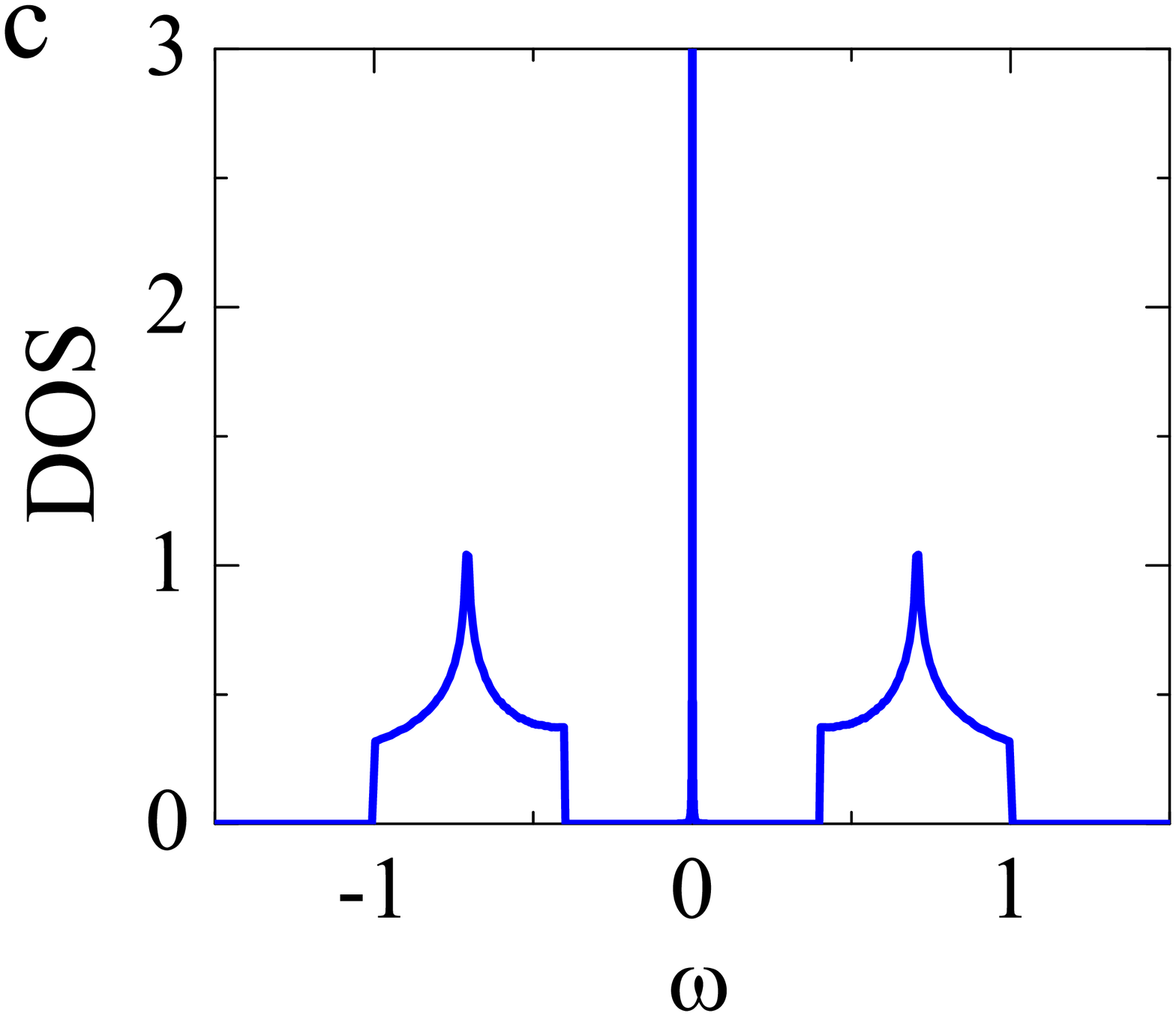} &
    \includegraphics[width=0.24\textwidth]{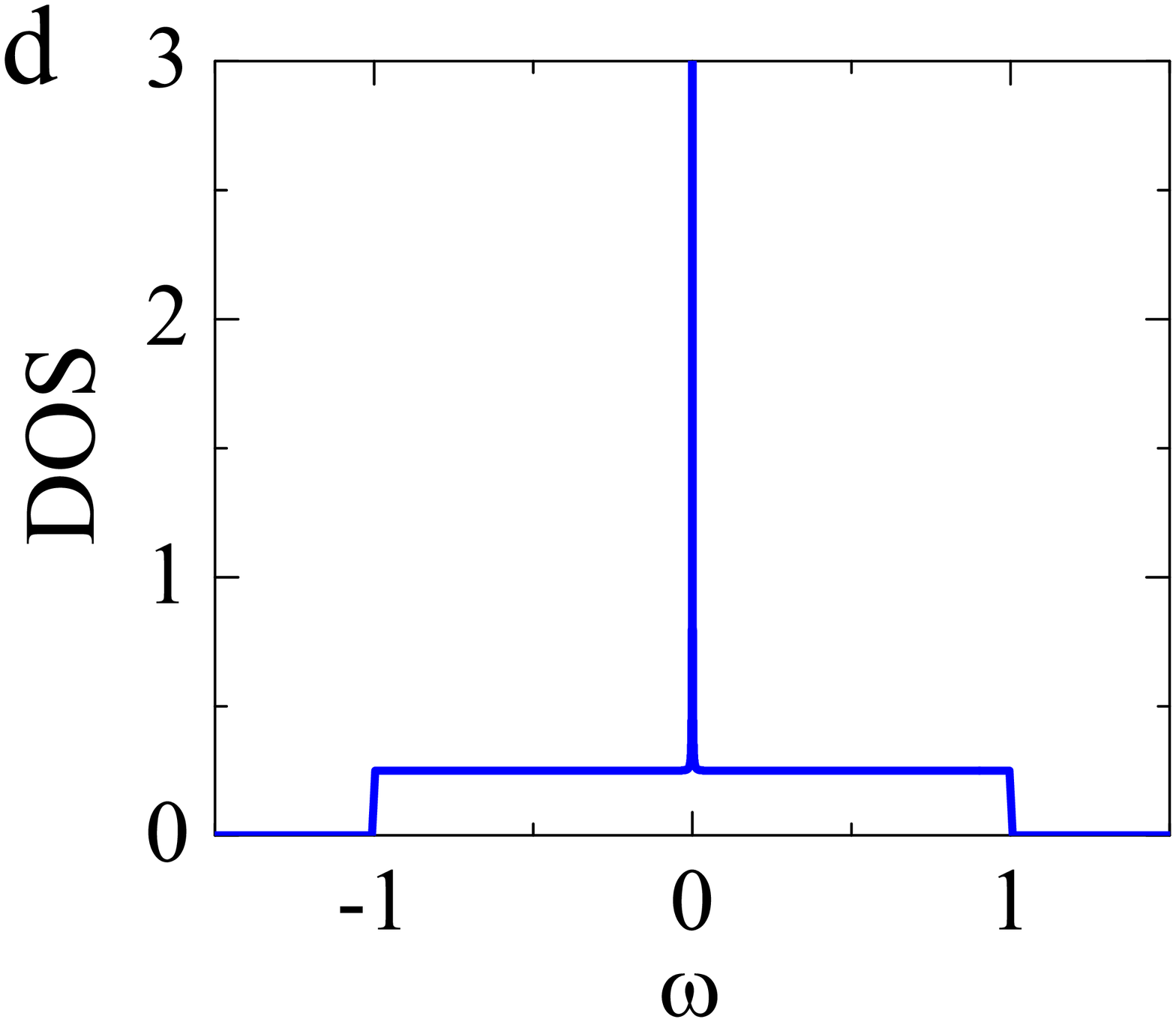}
  \end{tabular}
  \caption{(Color online) a, The Lieb lattice structure. b, The local DOS of the tight-binding model in the Lieb lattice at $C$ site (black dashed line) and at $A$ site (blue solid line). c, The local DOS of the tight-binding model with the intrinsic SOC in the Lieb lattice at $A$ site ($\lambda=0.1$). d, The DOS of the toy model in Eq. (\ref{toy}). The broadening parameter for these DOS $\eta=10^{-5}$.}
\label{fig1}
\end{figure}

A simplest model for flat-band lattices is the tight-binding model in the Lieb lattice \cite{Lieb,Tasakireview,Tien}.
The Lieb lattice is a square lattice with additional sites located in
the middle of every edge of the lattice squares (see Fig. \ref{fig1}). It is the basic structure of layered cuprates, and
has attracted research attention since the discovery of high temperature superconductivity \cite{Tasakireview}. Currently, the Lieb lattice can be realized by an array of optical waveguides \cite{Vicencio1,Vicencio2,Mukherjee}, loading ultracold atoms in optical lattices \cite{Taie}, and a molecular design \cite{Slot}.
A magnetic impurity or add-atom can be placed on the top of the lattice.
The Hamiltonian which describes a magnetic impurity (or a correlated add-atom) embedded in the Lieb lattice reads
\begin{eqnarray}
H &= & -t\sum_{<i,j>,\sigma} c^{\dagger}_{i\sigma} c_{j\sigma} +
\varepsilon_f \sum_{\sigma} f^\dagger_{o\sigma} f_{o\sigma} +
U n^f_{\uparrow} n^f_{\downarrow}  \nonumber \\
&& + V \sum_{\sigma} c^{\dagger}_{o\sigma} f_{o\sigma} + \text{H.c.}, \label{ham}
\end{eqnarray}
where $c^{\dagger}_{i\sigma}$ ($c_{i\sigma}$) is the
creation  (annihilation) operator of itinerant electron with spin $\sigma$ at lattice site $i$.
$t$ is the nearest neighbor hopping parameter. $f^{\dagger}_{o\sigma}$ ($f_{o\sigma}$) represents the creation (annihilation) operator of  the magnetic impurity at site $o$.  $U$ is the local Coulomb
interaction of electrons at the magnetic impurity. $n^f_\sigma=f^{\dagger}_{o\sigma}f_{o\sigma}$. $V$ is the hybridization strength between itinerant electrons and the magnetic impurity.
We are mainly interested in the case, where the impurity is placed at an edge center site ($A$ site in Fig. \ref{fig1}).
We also consider only the half filling case, where the chemical potential equals $0$.
Without the impurity, the local DOS at a corner site ($C$ site in Fig. \ref{fig1}) linearly vanishes at the Fermi energy like the Dirac electrons, whereas the one at a $A$ site exhibits the flat-band feature together with the Dirac electron pseudogap (see Fig. \ref{fig1}). With these features, the Lieb lattice allows us to study the Kondo effect in the opposite limits, where the DOS at the Fermi energy are either vanished or infinite.

The dynamics of the impurity can be determined by its Green function, which satisfies the Dyson equation
\begin{equation}
G_f(\omega) = \frac{1}{\omega-\varepsilon_f-\Gamma(\omega)-\Sigma_f(\omega)},
\end{equation}
where $\Sigma_f(\omega)$ is the self energy of the impurity, and
$\Gamma(\omega)=\Gamma g_c(\omega)/\pi$, $\Gamma=\pi |V|^2$. $g_c(\omega)$ is the bare local Green function of itinerant electrons at the impurity site.
The Kondo problem is completely determined by the hybridization function $\Gamma(\omega)$. The impurity contribution to a thermodynamical quantity $O$ is defined by
$
O_{\text{imp}} = O_{\text{tot}} - O_{\text{tot}}^{(0)},
$
where $O_{\text{tot}}$ is the thermodynamical quantity of the total system, and $O_{\text{tot}}^{(0)}$ is the one of the reference system without the impurity \cite{Bulla}.

We investigate the Kondo problem in the Lieb lattice by the NRG \cite{Wilson1,Wilson2,Wilson3,Bulla}. Within the NRG the hybridization function is discretized by a logarithmic mesh. However, in general the flat-band position falls out of the logarithmic mesh, and without modification the NRG cannot take into account the flat-band feature of the hybridization function. In order to overcome this failure, we broaden the hybridization function $\Gamma(\omega)=\Gamma\; g_c(\omega+i\eta)/\pi$ by a small parameter $\eta$.
Actually, we need only the imaginary part of $\Gamma(\omega)$ as the input information of conduction electrons in the NRG calculations.
The real part of $\Gamma(\omega)$ can be obtained from the imaginary part via the Kramer-Kronig tranformation.
We use the NRG Ljubljana package for our calculations \cite{Zitkoweb}. The energy unit is the half band width $D=2\sqrt{2}t=1$.
The hybridization function is discretized within the adaptive Z-scheme by the logarithmic mesh of the parameter $\Lambda=2$  \cite{Zitkoweb,Zitko1,Zitko2}. All calculated quantities are averaged over $Nz=8$ interleaved logarithmic meshes with the twist parameter $z=1/Nz$ \cite{Bulla,Zitkoweb,Zitko1,Zitko2}. The spectral functions are calculated by using full density matrix algorithm \cite{Zitkoweb,Delft}. The spin susceptibility is defined by
$T \chi = \langle (S^{z}_{\text{tot}})^2 \rangle - \langle S^z_{\text{tot}} \rangle^2$, where
$S^z_{\text{tot}}$ is the $z$-component of the total spin, and $T$ is temperature \cite{Bulla}.

\begin{figure}[b]
\vspace{-0.5cm}
\begin{center}
\includegraphics[width=0.45\textwidth]{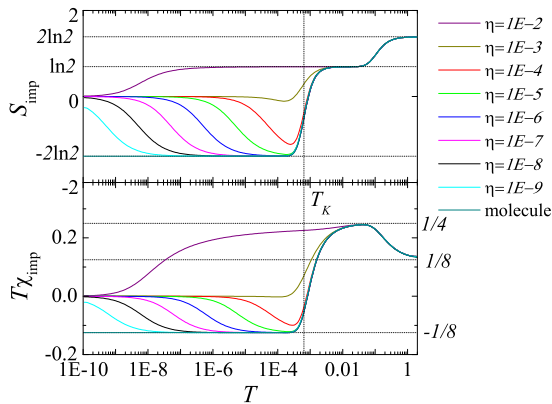}
\end{center}
\vspace{-0.5cm}
\caption{(Color online) The temperature dependence of the impurity entropy ($S_{\text{imp}}$) and the spin susceptibility ($T \chi_{\text{imp}}$) for different broadening parameters $\eta$ when the impurity is placed at $A$ site. The black-cyan solid lines are the results obtained by the exact diagonalization of the molecular model in Eq. (\ref{mol}). The vertical dotted line indicates the value of the molecular Kondo temperature $T_K$. The model parameters $\Gamma=0.001$, $U=0.5$, $\epsilon_f=-0.25$, $\epsilon_c=0$.}
\label{fig3}
\end{figure}

When the impurity is placed at a $C$ site, the hybridization function exhibits a linear pseudogap like the one in graphene or in Dirac electrons. This Kondo problem was well studied \cite{Bulla,Vojta}. Our calculations do well reproduce the previous results \cite{Bulla,Vojta}. We notice in this case the obtained results are insensitive to the small broadening parameter $\eta$ ($\eta < 10^{-3}$), since the pseudogap of the hybridization function is insensitive to small $\eta$.
The situation is quite different when the impurity is placed at a $A$ site, since in this case a flat-band feature exists in the hybridization function. In contrast to the $C$-site impurity case, the thermodynamical quantities are sensitive to the broadening parameter $\eta$, as they are shown in Fig. \ref{fig3}. When the broadening parameter $\eta$ is large, typically larger than a characteristic energy scale $T_K$ (for the model parameters in Fig. \ref{fig3}, $\eta > 10^{-4}$), the system behaves like the Kondo impurity in normal metals. Actually, widening by large $\eta$, the flat band becomes a normal metal.
There are three distinguished regimes. At low temperatures, the impurity entropy and the spin susceptibility exponentially approach to zero when $T\rightarrow 0$. This yields the strong coupling (SC) regime, where conduction electrons quench the magnetic moment of impurity, and they together form the many-body Kondo singlet state. When temperature increases, the system crosses to the local moment (LM) regime, where the impurity behaves like a magnetic moment. In this regime, the impurity entropy equals to $\ln 2$, and the spin susceptibility approaches $1/4$ \cite{Wilson1,Wilson2,Wilson3,Bulla}. At high temperatures, the impurity is effectively decoupled from conduction electrons, and it yields the free orbital (FO) regime, where the impurity entropy and the spin susceptibility equal to $2\ln2$ and $1/8$, respectively \cite{Wilson1,Wilson2,Wilson3,Bulla}. When the broadening parameter is small,  typically smaller than $T_K$, the impurity
thermodynamics exhibits different behaviors.
The characteristic energy scale $T_K$ separates the low and high temperature properties.
At high temperature $T > T_K$, the thermodynamical quantities are almost insensitive to the broadening parameter $\eta$. There are two distinguished regimes, the LM and the FO regimes, as they are shown in Fig. \ref{fig3}.
In the low temperature regime, the impurity thermodynamics
depends on the broadening parameter $\eta$.
At very low temperatures, typically smaller than $\eta$, the impurity entropy and the spin susceptibility also exponentially approach to zero when $T\rightarrow 0$. The finite value of $\eta$ would artificially make the flat band as a narrow band with a narrow Fermi sea. The electrons from this narrow Fermi sea still could quench the magnetic moment of impurity and they together also form the many-body Kondo singlet state. This  artificial SC regime always exists as long as $\eta$ is finite. However, it seems when $\eta \rightarrow 0$, the SC regime will disappear.
At higher temperatures, $\eta < T < T_K$, the impurity entropy and the spin susceptibility approach to constant values, which are insensitive to the small $\eta$. We call this regime the molecular SC. This constitutes a novel regime of the Kondo problem in the presence of band flatness.
As we will see, this regime is precisely the SC in a molecule of two orbitals, for instance a ligand orbital and a strongly correlated orbital. Two electrons are distributed over these two orbitals.
The Hamiltonian of the molecule reads \cite{Fuldebook}
\begin{eqnarray}
H_{\text{mol}} &=& \sum_{\sigma} (\varepsilon_c  c^\dagger_\sigma c_\sigma +
\varepsilon_f  f^\dagger_\sigma f_\sigma )+ U n^f_\uparrow n^f_\downarrow \nonumber \\
&& +
\sqrt{\frac{\Gamma}{2\pi}}  \sum_\sigma (c^\dagger_\sigma f_\sigma + \text{H.c.} ) , \label{mol}
\end{eqnarray}
where $c^\dagger_\sigma$ and $f^\dagger_\sigma$ represent the creation of an electron at the ligand and the strongly correlated orbitals, respectively. $\varepsilon_c$ and $\varepsilon_f$ are the orbital energy levels.
The ligand and the strongly correlated orbitals mimic  the conduction band and the magnetic impurity in bulk. The molecular model in Eq. (\ref{mol}) can analytically be analyzed in the strong correlation limit $U\rightarrow \infty$ \cite{Fuldebook}. When $\Gamma=0$, the ground state of the molecule is four-fold degenerate and is formed by one electron in the ligand orbital and the other electron in the strongly correlated orbital. One state is the spin singlet $|S\rangle$, and three others are the spin triplet $|T\rangle$ \cite{Fuldebook}
\begin{eqnarray*}
|S\rangle &=& \frac{1}{\sqrt{2}} (f^\dagger_{\uparrow} c^\dagger_\downarrow - f^\dagger_\downarrow  c^\dagger_{\uparrow})|0\rangle , \\
|T_{0}\rangle &=& \frac{1}{\sqrt{2}} (f^\dagger_{\uparrow} c^\dagger_\downarrow + f^\dagger_\downarrow  c^\dagger_{\uparrow})|0\rangle ,\\
|T_{\sigma}\rangle &=& f^\dagger_{\sigma} c^\dagger_\sigma |0\rangle ,
\end{eqnarray*}
where $|0\rangle$ is the vacuum state.
The excited states are $c^\dagger_{\uparrow} c^\dagger_\downarrow |0\rangle$, and $f^\dagger_{\uparrow} f^\dagger_\downarrow |0\rangle$. The latter excited state is excluded from the consideration
in the strong correlation limit. When the hybridization is turned on, the spin singlet couples with the ligand excited state $c^\dagger_{\uparrow} c^\dagger_\downarrow |0\rangle$, while the spin triplet remains unchanged. This coupling splits the singlet from the triplet, and separates the low-lying spin excitation from the charge one. The energy gain due to the spin singlet formation is
of order $\Gamma/\pi|\varepsilon_c-\varepsilon_f|$ \cite{Fuldebook}.
This is the SC regime of the Kondo problem in the molecule. The total entropy and spin susceptibility are zero like the ones of the Kondo effect in bulk. However, the impurity (or strongly correlated orbital) contributions to the entropy and spin susceptibility are not zero. They exactly equal to the opposite values of the entropy and spin susceptibility of the free ligand orbital in order to compensate the free ligand orbital contributions. Without the strongly correlated orbital, two electrons in the ligand orbital form a free orbital. Their entropy is $2\ln2$ and their spin susceptibility equals to $1/8$. Therefore, at low temperatures the impurity contribution to entropy is exactly $-2\ln2$, and the one to the spin susceptibility equals to $-1/8$, as one can see in Fig. \ref{fig3}.
When temperature increases, the triplet populates and forms a magnetic moment. At temperature $T > T_K$, the singlet-triplet splitting is unimportant.
The triplet states significantly contribute to the entropy and the spin susceptibility. As a consequence, with increasing temperature the system crosses to the LM regime, and then to the FO regime like the Kondo impurity in metals. The energy gain due to the spin singlet formation is the characteristic energy scale, which separates the SC regime from the other ones. It is called the molecular Kondo temperature \cite{Fuldebook}.
For finite $U$, we calculate the entropy and the spin susceptibility of the molecule by the exact diagonalization. Figure \ref{fig3} shows the impurity entropy and spin susceptibility of the molecule are identical to the ones in the Lieb lattice, except for the artificial SC regime due to the  artificially broadening of the flat band. This also indicates except for the artificial SC regime, the impurity thermodynamics in the lattice and the molecular models described by Eqs. (\ref{ham})-(\ref{mol}) are equivalent.
Therefore, the molecular Kondo temperature is also the characteristic energy scale $T_K$ of the Kondo problem in the Lieb lattice. It does not depend on the DOS of conduction electrons, and solely depends on the impurity parameters. In contrast to the many-body Kondo effect, where the Kondo temperature exponentially depends on the hybridization parameter, in the two-electron molecule in Eq. (\ref{mol}) the Kondo temperature has a power law of
the hybridization parameter.
In strong correlation regime, $T_K \approx \Gamma/\pi|\varepsilon_c-\varepsilon_f|$ \cite{Fuldebook}.
The obtained results also indicate
in the flat band lattices, only the flat band and the impurity are essential to the Kondo problem, and all dispersive bands are irrelevant.
In order to check whether any dispersive band is irrelevant to the Kondo problem or only the pseudogap bands in the Lieb lattice are irrelevant, we additionally
consider two cases. In the first case, the flat band is isolated from other dispersive bands by energy gaps. This case can be achieved by introducing the intrinsic spin-orbit coupling (SOC) to the Lieb lattice model \cite{Weeks}. This SOC is the spin and direction dependent hopping between next-nearest-neighbor $A$ sites. The SOC Hamiltonian can be written as
$
H_{\text{SOC}} = i \lambda \sum_{\langle\langle i, j\rangle\rangle,\sigma} \nu_{ij}
\sigma c^\dagger_{i\sigma} c_{j\sigma} ,
$
where $\lambda$ is the SOC strength, $\nu_{ij}=\pm 1$ depends on the clockwise (anticlockwise) hopping between $A$ sites \cite{Weeks}. The SOC in the Lieb lattice separates the flat band from two dispersive bands by the energy gaps, as one can see in Fig. \ref{fig1}.
In the second case, we study the Kondo problem in a toy model, where the DOS consists of a constant DOS and a Lorentzian narrow-band feature. The DOS in this toy model reads
\begin{equation}
\rho_c(\omega)=
\bigg\{\begin{array}{ll}
\frac{1}{\mathcal{N}}(\frac{1}{2D}+\frac{\eta}{\pi}\frac{1}{\omega^2+\eta^2}), &
\text{for} \;\;|\omega| \leq D \\
0, & \text{otherwise}
\end{array} , \label{toy}
\end{equation}
where $\mathcal{N}=1+2 \arctan(D/\eta)/\pi$ is the normalized factor of the DOS. The first term in the toy model DOS is the constant DOS, which represents conduction electrons of a normal metal. The second term represents a flat band with the Lorentzian broadening $\eta$. The DOS of this toy model is also plotted in Fig. \ref{fig1}.
We found that the impurity entropy and the spin susceptibility of these two models
are precisely identical to the ones of the original Lieb lattice described in Eq. (\ref{ham}).
This confirms in the presence of a flat band, only the flat band is relevant to the Kondo problem, and all dispersive bands are irrelevant. The dispersive band quenching is independent on the position and the dispersion relation of the bands.
One may expect in the limit $\eta \rightarrow 0$, the artificial SC regime will disappear, and only the molecular SC is stable at low temperature.
Actually, in the limit $\eta \rightarrow 0$, the effective hybridization is infinite, hence the impurity exclusively couples with the electron at the $A$ site. In the SC regime, they form the spin singlet, and it is purely local.
This reduces the many-body Kondo problem to the two-electron molecular one.
The molecular Kondo effect still occurs for finite $\eta$, providing $\eta< T < T_K$.
By tuning the ratio $\eta / T_K$, one would observe a crossover from the many-body Kondo effect to the molecular one, as one can see in Fig. \ref{fig3}. In this crossover, the Kondo temperature changes from the exponential to power laws of the hybridization strength. The molecular Kondo temperature usually is bigger than the many-body Kondo temperature, as one can see in Fig. \ref{fig3}.

\begin{figure}[t]
\vspace{-0.5cm}
\includegraphics[width=0.4\textwidth]{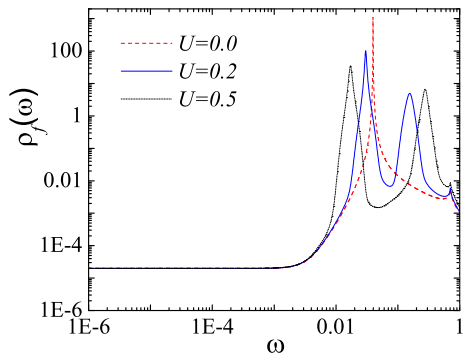}
\caption{(Color online) The spectral function of impurity for different values of $U$ for the symmetric case $\epsilon_f=-U/2$. ($\Gamma=0.01$, $T=10^{-5}$).}
\label{fig5}
\end{figure}

The scattering in the Lieb lattice due to the impurity is determined by the $T$-matrix, which essentially relates to the impurity spectral function. The quasiparticle interference also relates to the impurity spectral function. In Fig. \ref{fig5} we plot the spectral function of impurity in the molecular SC regime for the particle-hole symmetric case $\varepsilon_f=-U/2$. We have used the Gaussian broadening parameter $\delta=0.2$ for the spectral function, since in the NRG the spectral function is merely discrete energy points with spectral weights \cite{Bulla}.
When $U=0$, the spectral function exhibits only two symmetric peaks in the positive and negative frequency domains. These peaks reflect the charge excitation in the noninteracting case. For finite $U$,
in contrast to the Kondo effect in metals, the spectral function of impurity does not exhibit the Kondo resonance at the Fermi energy. Instead of the Kondo resonance, two narrow peaks and two broader peaks appear on the incoherent background of the dispersive bands of the Lieb lattice. The narrow peaks are the spin excitation and the broader peaks are the charge excitation as the ones in the molecule \cite{Fuldebook}. The separation of the spin and charge excitations is essentially an electron correlation effect. This separation is proportional to $U/2$.

\section{Conclusion}

We have solved the Kondo problem in the Lieb lattice, where a band flatness is present.
We have found a novel regime, the molecular SC, of the Kondo problem in flat-band and narrow-band systems. The band flatness prevents the participation of dispersive electrons in the Kondo singlet formation and reduces the many-body Kondo problem to the two-electron one. The impurity contributions to the thermodynamical and dynamical quantities in the molecular Kondo effect are qualitatively different in comparison with the ones in the many-body Kondo effect. The molecular Kondo effect can be observed in any narrow-band system, providing $\eta < T < T_K$, where $\eta$ is the half width of the narrow band, and $T_K$ is the Kondo temperature of the two-electron two-orbital molecule. The molecular Kondo temperature is solely dependent on molecular parameters, and in contrast to the many-body Kondo effect, it is independent on the DOS of conduction electrons.
By varying the ratio $\eta/T_K$, we could drive the system from the many-body Kondo effect to the molecular one. In this crossover the characteristic energy scale of the system changes from the exponential to power laws of the hybridization strength.
This hints a possibility for experimental observation of the molecular Kondo effect or  an entanglement of two spin qubits in materials.

\section*{Acknowledgement}

One of the authors (M.T.) acknowledges the ICTP Asian Network in Condensed Matter and Complex Systems, and the hospitality of the Center for Theoretical Physics of Complex Systems at Daejeon.
This research is funded by Vietnam National Foundation
for Science and Technology Development (NAFOSTED) under Grant No 103.01-2017.13.

\end{document}